# Optical Properties of the South Pole Ice
# at Depths Between 0.8 and 1 km

P. Askebjer,* S.W. Barwick,† L. Bergström,* A. Bouchta,* S. Carius,‡ A. Coulthard,§
K. Engel,§ B. Erlandsson,* A. Goobar,* L. Gray,§ A. Hallgren,‡ F. Halzen,§
P.O. Hulth,* J. Jacobsen,§ S. Johansson,*,¶ V. Kandhadai,§ I. Liubarsky,§ D. Lowder,∥
T. Miller,∥,** P.C. Mock,† R. Morse,§ R. Porrata,† P.B. Price,∥ A. Richards,∥
H. Rubinstein,‡ E. Schneider,† Q. Sun,* S. Tilav,§ C. Walck* & G. Yodh†

The optical properties of the ice at the geographical South Pole have been investi-
gated at depths between 0.8 and 1 kilometers. The absorption and scattering lengths
of visible light ($\sim$515 nm) have been measured *in situ* using the laser calibration setup
of the AMANDA neutrino detector. The ice is intrinsically extremely transparent. The
measured absorption length is $59 \pm 3$ meters, comparable with the quality of the ultra-
pure water used in the IMB and Kamiokande proton-decay and neutrino experiments
and more than two times longer than the best value reported for laboratory ice. Due to a
residual density of air bubbles at these depths, the trajectories of photons in the medium
are randomized. Assuming bubbles are smooth and spherical, the average distance be-
tween collisions at 1 km depth is about 25 cm. The measured inverse scattering length
on bubbles decreases linearly with increasing depth in the volume of ice investigated.

The AMANDA (Antarctic Muon And Neutrino Detector Array) project was conceived to exploit
polar ice as a transparent and sterile detection medium to attain large detection volumes for muons
and neutrinos from astrophysical sources. Photomultiplier tubes (PMTs) deployed in the South Pole

* *Stockholm University, Sweden*
† *University of California, Irvine, USA*
‡ *Uppsala University, Sweden*
§ *University of Wisconsin, Madison, USA*
¶ *Currently at Jönköping University, Sweden*
∥ *University of California, Berkeley, USA*
** *Currently at Bartol Research Institute, Delaware, USA*

ice sense the Cherenkov light emitted by highly relativistic muons. Downgoing muons originate from cosmic-ray showers, and nearly isotropic muons result from interactions between neutrinos and nucleons in the ice and rock around the detector. The polar ice, unlike ocean water which has also been proposed as a neutrino detector medium (*e.g.*, the DUMAND and NESTOR experiments [1, 2]), is free of bioluminescent organisms and natural radioactive isotopes such as $^{40}$K. In addition, the ice forms a rigid support structure for the detector elements. At the low temperature of the medium, -55 ° C, the measured thermal noise rates average below 2 kHz per PMT.

One of the characteristic features of shallow polar ice is the abundance of air bubbles [3]. These are formed as pockets of air get trapped between grains of snow [4]. Microscopic examinations of ice-core samples from Greenland and Antarctica [5, 6] show that, at first, the size of air bubbles decreases with increasing depth (*i.e.* pressure). When the hydrostatic pressure exceeds the formation pressure of air hydrates, a transition occurs where the number of bubbles decreases. The hydrate crystals that form have a refractive index only 0.4 % larger than that of pure ice [7], making scattering on these crystals negligible [8]. At the Russian Antarctic station of Vostok, where the temperature of the ice is similar to that of the South Pole ice (the accumulation rate of snow and the altitude are, however, different) bubbles were *not* seen in ice cores below 1280 m depth, setting an upper limit to the number density of air bubbles of 0.5 cm$^{-3}$ [6]. In the ice cores extracted from the Antarctic Byrd station, bubble-free ice was found below 1100 meters depth [4].

Previous measurements of the flux of Cherenkov photons from down-going cosmic-ray muons at 800 meters depth with a prototype system consisting of four small PMTs (7.5 cm diameter) on a single string were tentatively interpreted to indicate that the South Pole ice was bubble-free at that depth [9]. In this study, employing many more PMTs of larger size and a laser calibration system, we have now, however, found clear indications of a residual density of air bubbles between 0.8 and 1 km, giving a very short *scattering* length of the Cherenkov light. The *absorption* of photons, however, is remarkably low. This makes the site, at deeper levels, essentially free from air bubbles, very promising for the AMANDA detector.

During the Antarctic summer 1993-94, four strings, each equipped with twenty PMTs (20 cm diameter), were installed between 800 and 1000 meters depth in the South Pole ice as shown in figure 1. Along with the main signal cable carrying the high voltage and the PMT signals, optical fibers carry light from a laser calibration source at the surface to each optical module. Each optical fiber terminates in a diffusive nylon sphere placed about 30 cm from the PMT. By sending laser pulses to individual nylon spheres and measuring the photon travel-time distributions between modules (time resolution $\sim 5$ ns), the optical properties of the medium can be derived. The intensities of the laser pulses coming out of the spheres are high enough that photons are detected by PMTs in neighboring strings. The calibration pulses have a wavelength of $515 \pm 15$ nm . The travel times of photons between the emitting nylon sphere and the receiving optical modules were found to be many times longer than what would be expected from a straight path. Figure 2 shows two examples of such measurements. The geometrical distance between the source and the detector is 21 and 32 meters in figures 2a and 2b, respectively. With no obstacles, the signals would have arrived after 91 ns and 142 ns, respectively.

A model for the distribution of the photon travel times can be constructed, assuming that the time delays are caused by collisions with air bubbles. The scattering length on spherical bubbles, *i.e.* the average distance that a photon travels between collisions with bubbles, is

$$\lambda_{bub} = \frac{1}{n_{bub} \langle \pi r^2 \rangle_{bub}}, \tag{1}$$

where $n_{bub}$ is the density of bubbles and $\langle \pi r^2 \rangle_{bub}$ their average geometrical cross section. Dust particles, although present in glacier ice with a comparable number density, have typically 10 to 100 times smaller radii than air bubbles, *i.e.* the collision rate between photons and insoluble dust particles can be neglected in the presence of air bubbles [8].

If the light is detected at the PMT after a large number of scatterings the situation can be described as a random walk (diffusive) process. We only have to incorporate two modifications to the standard random walk treatments given in the literature (*e.g.*, [10]), namely absorption of light and the non-isotropy of the scattering amplitude for photons incident on air bubbles.

Absorption can be taken into account by the factor $e^{\frac{-N\lambda_{bub}}{\lambda_a}}$ for each path of $N$ steps (we assume $N >> 1$; typically $N$ is on the order of 1000 in our problem), where $\lambda_a$ is the absorption length.

Non-isotropy of the scattering means that there is a correlation between successive vectors making up the random walk. If the bubbles are spherical there is still an azimuthal symmetry, and one can show that (*e.g.* by introducing rotation matrices which rotate each successive vector to the polar axis)

$$\langle \vec{r}_1 \cdot \vec{r}_{1+k} \rangle = \lambda_{bub}^2 (\delta_{0,k} + \tau^k), \tag{2}$$

where $\tau = \langle \cos(\theta) \rangle$ is the average of the cosine of the scattering angle.

In the formula for random walk with absorption [11]

$$W_N(\vec{R}) = \frac{1}{(2\pi \langle \vec{R}^2 \rangle_N / 3)^{3/2}} e^{-\frac{3\vec{R}^2}{2\langle \vec{R}^2 \rangle_N}} e^{\frac{-N\lambda_{bub}}{\lambda_a}}, \tag{3}$$

$W_N(\vec{R})$ is the probability function for reaching $\vec{R}$ after $N$ steps and

$$\langle \vec{R}^2 \rangle_N = \langle (\sum_{j=1}^N \vec{r}_j)^2 \rangle$$
$$\propto 2N\lambda_{bub}^2 (1 + \tau + \tau^2 + \tau^3 + .. + \mathcal{O}(1/N)) = \frac{2N\lambda_{bub}^2}{1 - \tau}. \tag{4}$$

With the identification $N\lambda_{bub} = c_i t$, where $c_i = c/n$ is the velocity of light in the ice, we obtain the formula corresponding to the Green's function for the radiative transport of a spherically symmetric light pulse emitted at a distance $d = 0$ at time $t = 0$

$$u(d,t) = \frac{1}{(4\pi Dt)^{3/2}} e^{\frac{-d^2}{4Dt}} e^{\frac{-c_i t}{\lambda_a}}, \tag{5}$$

where $u(d,t)$ is the density of photons (normalized to unity at $t=0$) at a distance $d$ from the source at time $t$ and the constant of diffusion $D$ is given by

$$D = \frac{c_i \lambda_{eff}}{3}, \tag{6}$$

with the effective scattering length $\lambda_{eff}$ related to $\lambda_{bub}$ through the formula

$$\lambda_{eff} = \frac{\lambda_{bub}}{1 - \tau}. \tag{7}$$

We note that in the expression for $D$ only the refractive scattering part of the optical amplitude has to be taken into account. If diffraction is included, making the effective bubble cross section increase from $\sigma$ to $(B + 1)\sigma$ [12], then $\lambda_{bub} \to \lambda_{bub}/(1 + B)$, $\tau \to (B + \tau)/(B + 1)$, and $D$ is easily seen to be invariant under these transformations.

The distributions of arrival times were also calculated in a Monte Carlo simulation. Thanks to the existence of widely separated length scales, $d \sim \lambda_a >> \langle r_{bub} \rangle >> \lambda_L$, where $\lambda_L$ is the wavelength of the laser light, a fairly simple optical treatment is adequate for this analysis. In fact, the estimate of $\lambda_a$ is obtained almost directly from the fall-off of the arrival times of late photons, and is fairly insensitive to the uncertainties in the determination of $\lambda_{bub}$.

The initial state in any single scattering can be regarded as being completely incoherent. We have thus employed simple geometrical optics for the scattering process incorporating singly refracted light, externally reflected light, and totally reflected light with relative strengths of the components given by the Fresnel coefficients. We take into account the change of refraction index of the air in the bubbles assuming hydrostatic pressure. For the ice we use the value of the refraction index, $n = 1.31$, given in [13]. With our approximations we find for *spherical, smooth* bubbles (as observed in ice core samples from the Antarctic Byrd station [4]) that $\tau$ in Eq. (7) takes a value around 0.75 (corresponding to an average scattering angle of 41°).

In figure 2 we show a comparison of the experimental data, the results of our Monte Carlo simulations, and the analytical formula of Eq. (5). The PMTs considered are at 21 and 32 meters distance from the emitting nylon sphere in figures (a) and (b) respectively. The best fit to the data gives $\lambda_{bub} = 0.19$ m in both cases, and $\lambda_a = 61$ m in (a) and $\lambda_a = 60$ m in (b). The agreement between the Monte Carlo simulations (run with the above parameters), the data and the analytical fit is quite remarkable. We thus conclude that the physics behind the PMT calibration data is well understood, and that we have at our disposal a tool to measure properties of the ice such as $\lambda_a$ and $\lambda_{eff}$. These quantities are uniquely determined by the temporal distribution of photons at each particular PMT, once the geometry of the detector array is fixed.

A global $\chi^2$ minimization of the time data was performed where the relative vertical distances between the strings were left as free parameters, $\lambda_a$ was assumed to be a constant and $1/\lambda_{bub}$ was expressed as a linear function of depth. Although all the optical modules are equipped with an optical fiber and a nylon sphere, only 12 of them could be pulsed before the members of the AMANDA experiment had to leave the site in February 1994. Further measurements, at several wavelengths, will be performed in the future. For the present analysis we have selected time distributions where at least 1000 pulses were registered by the receiving PMT. The time measured is set by the arrival of the *first* photon. In order to reproduce the true time distribution given by the different photon paths in the medium, we selected distributions where the probability for more than one photon detected at the receiving PMT for each trigger is less than 6 %. A total of 36 time distributions like the ones shown in figure 2 were used in the global fit.

The optical properties of the medium are parameterized as described in Eq. (5). Only statistical uncertainties in the number of detected photons were considered. The relative vertical positions of the four strings were determined to about 0.2 meter precision. For the absorption length and the inverse scattering length on air inclusions, the global fit yields:

$$\lambda_a = 59 \pm 1 \text{ m, and}$$

$$\frac{1}{\lambda_{bub}} = \frac{(28.6 \pm 1.1) - (0.025 \pm 0.001)z}{4(1-\tau)} \ m^{-1}, \tag{8}$$

where $z$ is the vertical depth (in meters), defined to be zero at the surface and increasing downwards. We have used Monte Carlo simulations to estimate the systematic uncertainties in the global fit from the contamination of multiple-photon events, background events, and limited sample sizes. We find that the various effects tend to cancel each other. A conservative estimate of the total uncertainty in $\lambda_a$ is:

$$\lambda_a = 59 \pm 1(stat) \pm 3(syst) \text{ meters.}$$

Figure 3(a) shows the observed linear dependence of $\frac{1}{\lambda_{bub}}$ on depth. This depth dependence is steeper than would be expected if the bubble number density were constant and only the bubble sizes decreased under the hydrostatic pressure ($\frac{1}{\lambda_{bub}} \propto z^{-\frac{2}{3}}$). The shaded area in Fig. 3(b) corresponds to the uncertainty in the bubble shapes, *i.e.* in the average scattering angle. Also shown in the figure are results from the measurements on the Vostok and Byrd ice cores. Figure 4 shows the measured absorption lengths at different depths. The solid line is the fitted constant value.

The propagation parameters of visible light in the South Pole ice have been measured to approximately 5 % accuracy, using the calibration setup of the AMANDA detector. This is the first time that such measurements have been made without extracting ice samples. The small uncertainty in the measured value of $\lambda_a$ was only achievable because of the long travel times of photons in the bubbly ice. Alternative explanations to the observed photon time distributions such as these being caused by fluorescence in the medium are extremely unlikely. Our results are also incompatible with a localized bubble concentration as the cause of the observed time smearing. The examined ice volume, 0.8–1 km below surface, has an extremely long absorption length, comparable with the quality of the ultra-pure water used in the IMB and Kamiokande proton-decay and neutrino experiments [14, 15] and more than two times longer than the best value reported for laboratory ice [13].

The results of this study suggests that the ice cap is indeed an ideal medium for a neutrino telescope. If the absorption length does not deteriorate with depth, the volume of a future muon and neutrino detector to be deployed at greater depth, can be made significantly larger than previously anticipated, since the PMTs can be spaced further apart. A linear extrapolation of our data would yield that bubbles vanish at ∼1150 meters at the South Pole. The data from Vostok and Byrd (Fig 3(b)) show, however, that the rate of bubble disappearance becomes somewhat slower towards the end of the transition process. During the Antarctic summer 1994-95 we plan to collect new calibration data with the same system to establish the wavelength dependence of the optical properties of the South Pole ice.

# Figure Captions

Figure1:

The AMANDA detector. The four strings were deployed between December 1993 and January 1994. A 60 cm wide hole in the ice was drilled for each string by pumping water at 90°C at a rate of 40 gallons per minute. The time to drill a hole was less than 90 hours, typically 4 days. The holes are vertical within 1 mrad precision. There are 20 modules in each string, spaced by 10 meters. Each PMT is contained inside a pressure vessel. An optical fiber carries laser light from the ice surface to a diffusive nylon sphere suspended below each PMT.

Figure 2:

Arrival time distributions for two different source-detector separations. The data points (dots) are compared to the Monte Carlo simulation (histogram) and the fits of Eq. (5) (solid line). The emitting and receiving PMTs are 21 and 32 meters apart in figures (a) and (b) respectively.

Figure 3:

Inverse scattering length $\frac{1}{\lambda_{bub}}$ as a function of depth, $z$. The data points in (a) are AMANDA measurements shown together with the best fitted linear dependence of Eq. (8) for $\tau = 0.75$ (*i.e.* spherical bubbles). The fitted line extrapolates to zero at $\sim 1150$ meters. The shaded area in (b) shows the AMANDA results allowing for $\tau = 0 - 0.75$ ($\tau = 0$ corresponds to isotropic scattering angle). Also shown are the Byrd and Vostok ice-core measurements [4, 6] where we have computed an equivalent value of $\lambda_{bub}$ from their quoted bubble number density ($\mathcal{N}$) and average bubble radius squared ($\varrho^2$), $\frac{1}{\lambda_{bub}} = \mathcal{N} \pi \varrho^2$.

Figure 4:

Measured absorption length, $\lambda_a$, between 0.8 and 1 km depth. The solid line shows the best fitted value.

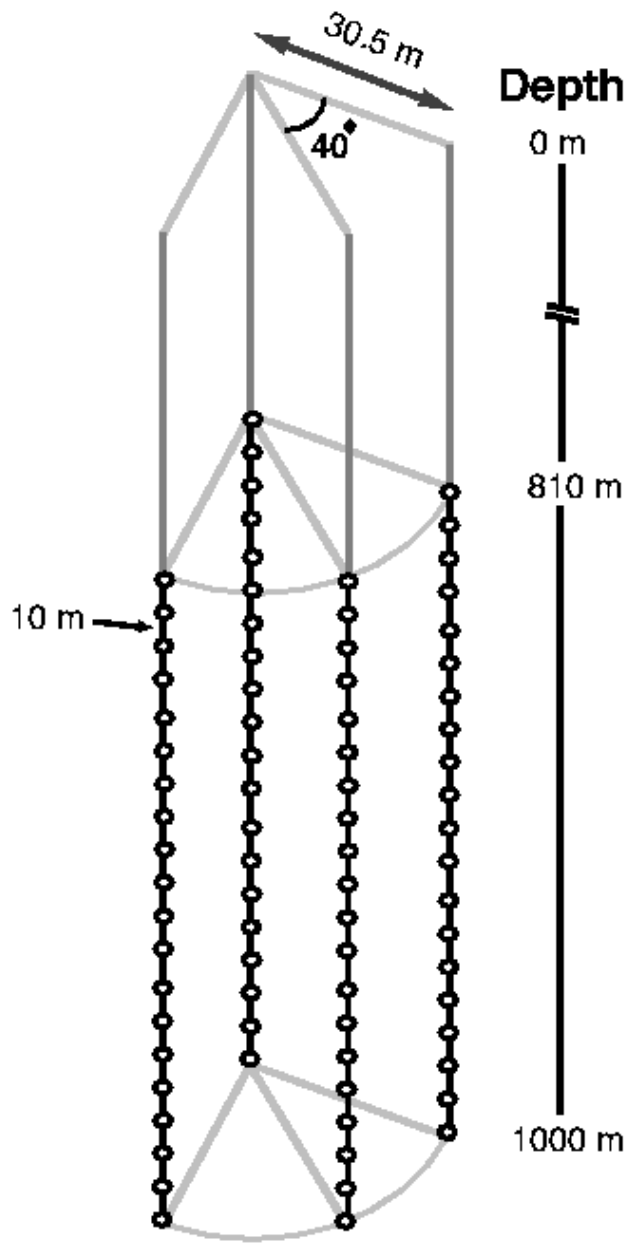

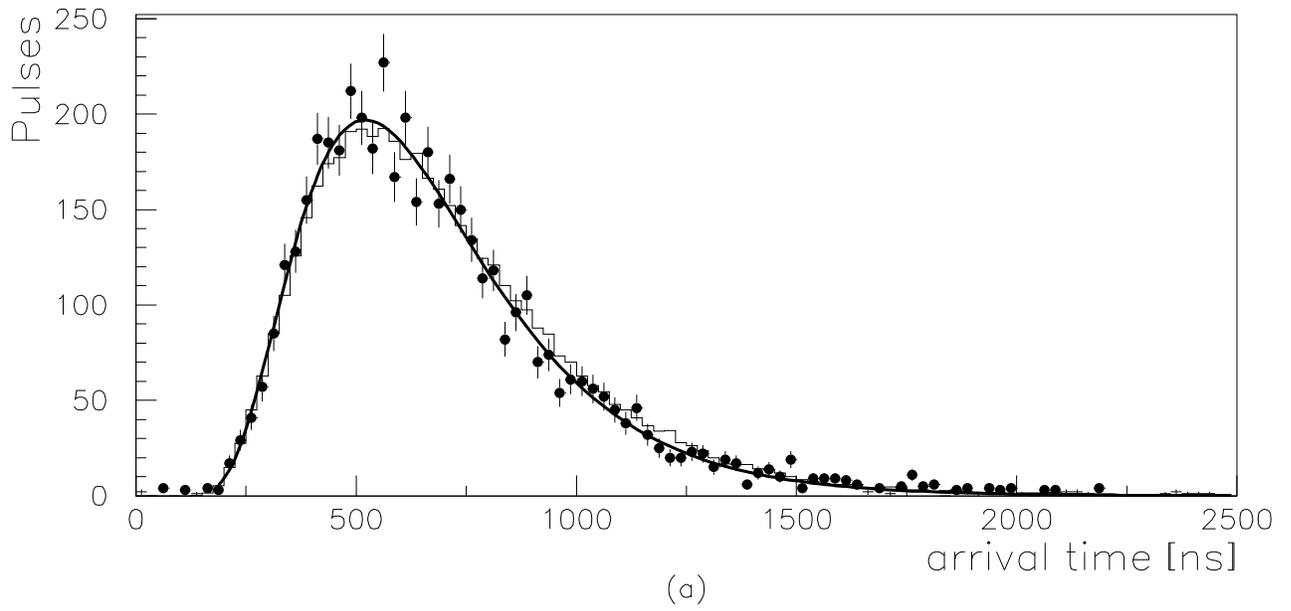

(a)

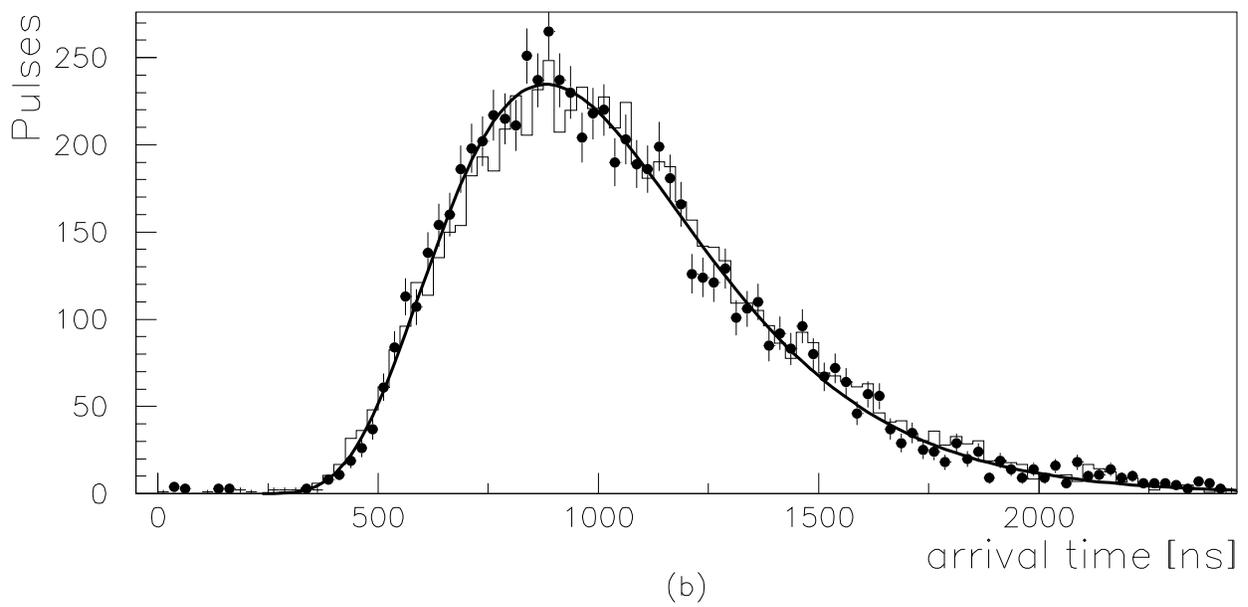

(b)

Figure 2:

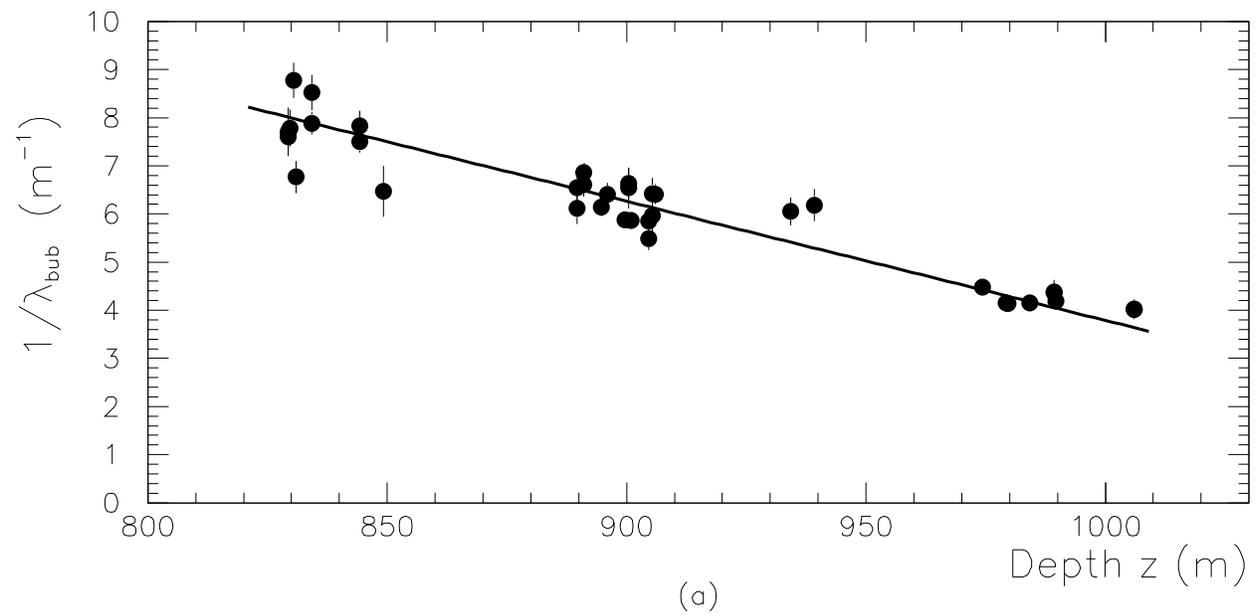

(a)

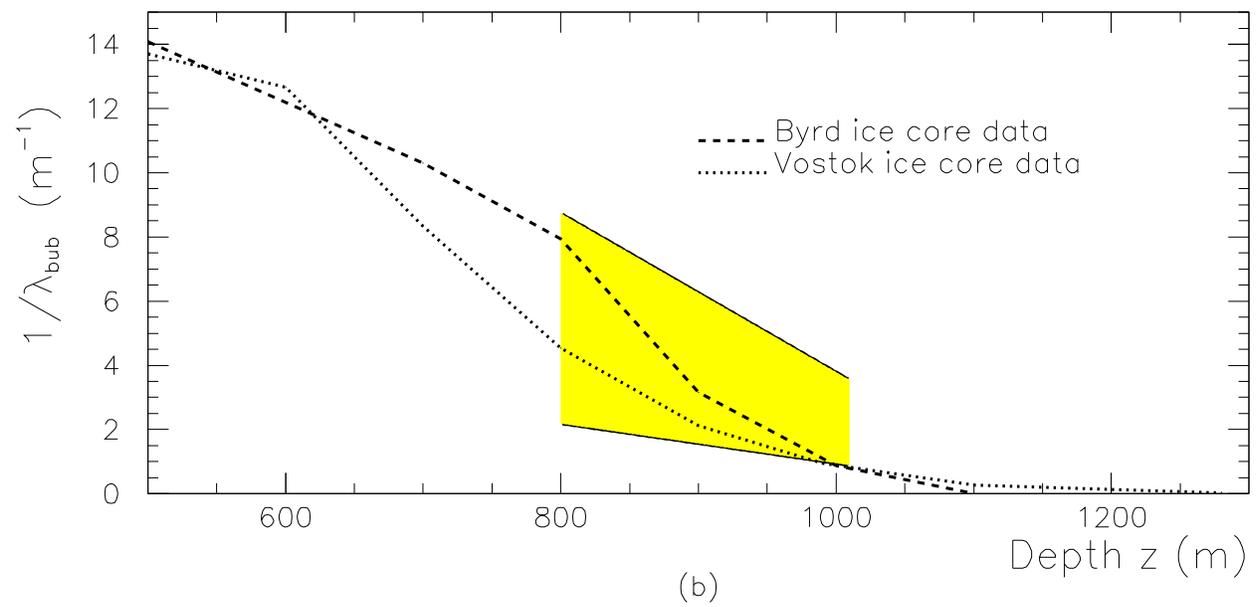

(b)

Figure 3:

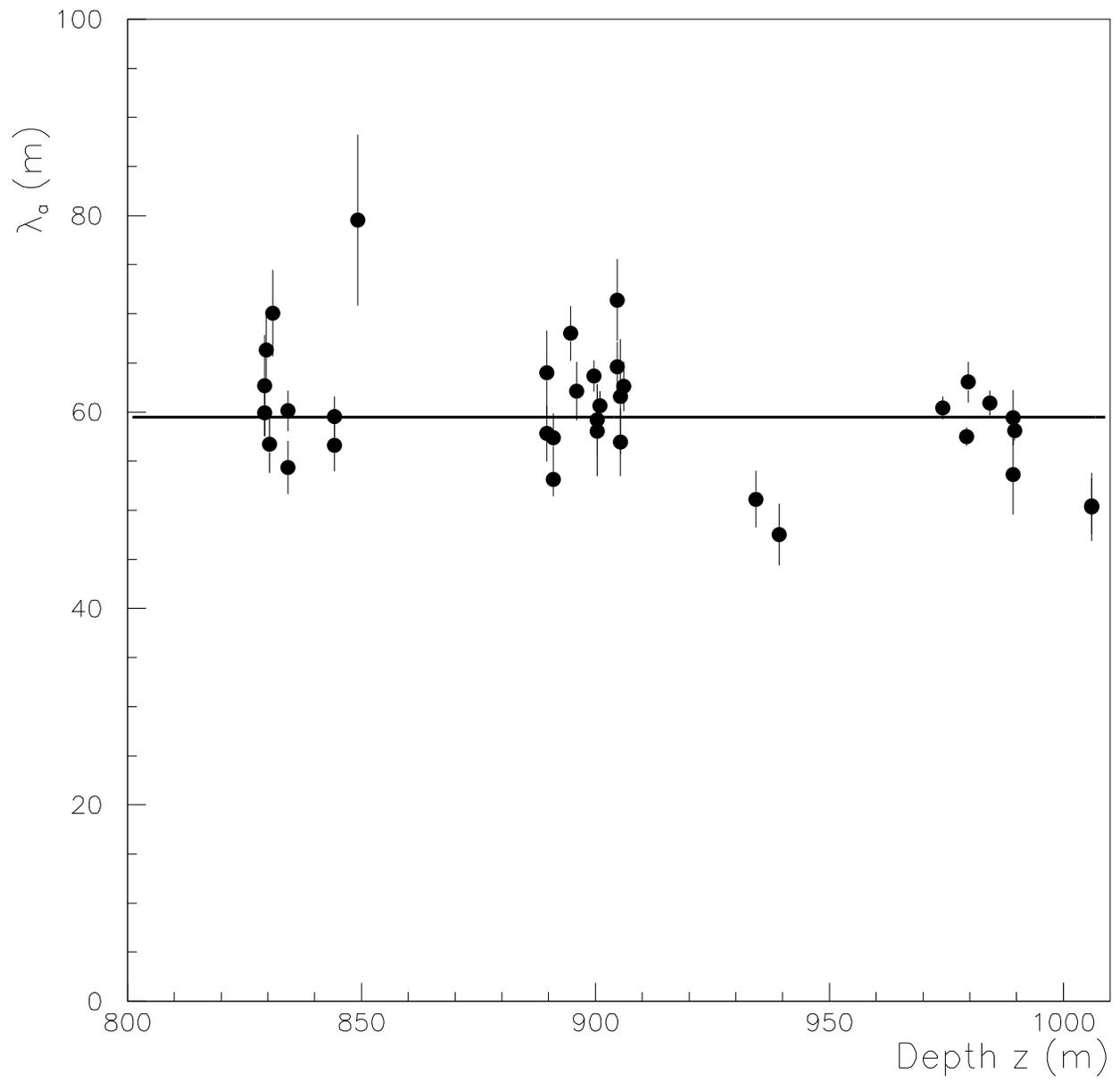

Figure 4: